\documentclass[aps,prb,showpacs,twocolumn]{revtex4-1}
\usepackage{amssymb}
\usepackage{amsmath}
\usepackage{graphicx}
\usepackage{dsfont}
\usepackage{times}

\begin{document}

\title{Scaling of the Reduced Energy Spectrum of Random Matrix Ensemble}
\author{Wen-Jia Rao}
\email{Corresponding author. wjrao@hdu.edu.cn}
\author{M. N. Chen}
\affiliation{School of Science, Hangzhou Dianzi University, Hangzhou 310027, China.}
\date{\today }

\begin{abstract}
We study the reduced energy spectrum $\{E_{i}^{(n)}\}$, which is constructed
by picking one level from every $n$ levels of the original spectrum $%
\{E_{i}\}$, in a Gaussian ensemble of random matrix with Dyson index $%
\beta\in \left( 0,\infty \right) $. It is shown the joint probability
distribution of $\{E_{i}^{(n)}\}$ bears the same form as $\{E_{i}\}$ with a
rescaled parameter $\gamma =\frac{n(n+1)}{2}\beta +n-1$. Notably, the $n$-th
order level spacing and gap ratio in $\{E_{i}\}$ become the lowest-order
ones in $\{E_{i}^{(n)}\}$, which explains their distributions found
separately by recent studies in a consistent way. Our results also establish
the higher-order spacing distributions in random matrix ensembles beyond
GOE,GUE,GSE, and reveal a hierarchy of structures hidden in the energy
spectrum.
\end{abstract}

\maketitle

\section{Introduction}

\label{sec1}

Random matrix theory (RMT) is a powerful mathematical tool for studying
complex quantum systems, it describes the universal properties of random
matrix that are determined only by the system's symmetry while independent
of microscopic details. For this reason, RMT has been applied to various
fields ranging from disordered nuclei to isolated quantum many-body systems%
\cite{Porter,RMP,PR}.

Among various statistical quantities of RMT, the most widely-used ones are
the distribution of nearest level spacings $\left\{
s_{i}=E_{i+1}-E_{i}\right\} $ and gap ratios $\left\{
r_{i}=s_{i+1}/s_{i}\right\} $. It is well established that in a chaotic
system, the distribution of level spacing $P\left( s\right) $ will follow a
Wigner-Dyson distribution\cite{Mehta,Haake2001} (see Eq.~(\ref{equ:WD}) in
Sec.~\ref{sec3}), which reveals the level repulsion in a direct way.
However, when accounting $P\left( s\right) $, an unfolding procedure is
required to erase the model-dependent information about local density of
states (DOS). In contrast, gap ratios distribution $P\left( r\right) $ is
independent of DOS and requires no unfolding procedure\cite{Oganesyan,Atas},
and has found various applications especially in the context of many-body
localization (MBL)\cite%
{Oganesyan,Avishai2002,Regnault16,Regnault162,Huse1,Huse2,Huse3,Sarma,Luitz,Rao182,Sarkar,Corps}%
.

Both the nearest level spacing and gap ratio account for the short-range
level correlations. However, long-range correlations are also important,
especially in the study of MBL transition phenomena. Actually, there're a
number of RMT models accounting for the intermediate level statistics
between Wigner-Dyson and Poisson ensembles\cite{Data,Jain,Auberson,SRPM,Ndawana,Mirlin}, and some of them are suggested
to describe the level statistics in MBL transition regime\cite%
{Shukla,Serbyn,Sierant19,Buijsman,Sierant20}, all of which more or less
describe the short-range level correlations well, while their difference can
only be revealed when long-range correlations are concerned. Commonly, the
long-range correlation in a random matrix ensemble is described by the
number variance $\Sigma ^{2}$ or the Dyson-Mehta $\Delta _{3}$ statistics%
\cite{Haake2001}, however, both of them are very sensitive to the concrete
unfolding strategy which has already been a source of misleading signatures%
\cite{Gomez2002}. Instead, it's more direct and numerically easier to study
the higher order level spacings and gap ratios, as have been done in a
number of recent works\cite%
{Sierant19,Sierant20,Tekur1,Tekur,Bhosale,Atas2,Chavda,Magd,Duras,Rubah,Rao20,Gong}%
.

Formally, the $n$-th order level spacing and gap ratio are defined as $%
\left\{ s_{i}^{\left( n\right) }=E_{i+n}-E_{i}\right\} $ and $\left\{
r_{i}^{\left( n\right) }=s_{i+n}^{\left( n\right) }/s_{i}^{\left( n\right)
}\right\} $ respectively. The higher-order gap ratios are first studied in
Ref.~[\onlinecite{Tekur}], where the authors provide strong numerical
evidences that $P\left( r^{\left( n\right) }\right) $ in a random matrix
ensemble with Dyson index $\beta $ bears the same form as $P\left( r\right) $
with a rescaled parameter
\begin{equation}
\gamma =\frac{n(n+1)}{2}\beta +n-1\text{,}  \label{equ:0}
\end{equation}%
where $\beta =1,2,4$ correspond to the Gaussian orthogonal ensemble (GOE),
Gaussian unitary ensemble (GUE) and Gaussian symplectic ensemble (GSE)
respectively. On the other hand, it is later proved in Ref.~[\onlinecite{Rao20}]
that $P\left( s^{\left( n\right) }\right) $ has the same form as
Wigner-Dyson distribution with a rescaled parameter $\gamma $ that is
identical to the one in Eq.~(\ref{equ:0}) for $\beta \in \left( 0,\infty
\right) $. This strongly hints a homogeneous relation between the
higher-order level spacing and gap ratio, but an explanation is still
lacking, which gives the first motivation of this work.

The second motivation comes from the recent works that encounter physical
systems that go beyond the three standard ensembles with $\beta =1,2,4$. For
example, the $\beta =3$ behavior has been found in a 2D lattice with
non-Hermitian disorder\cite{Tzortzakakis}; and the ensembles with
non-integer $\beta $ is suggested to describe the level statistics in the
whole region along the MBL transition in 1D random spin system\cite{Buijsman}%
, while their efficiency in describing long-range spectral correlations is
controversial\cite{Sierant20}. Therefore, it's beneficial to have an
expression for the higher-order spacing distributions in these ensembles,
which will also be offered in this study.

In this work, we find that the key to link the higher-order level spacing
and gap ratio is the reduced energy spectrum $\left\{ E_{i}^{\left( n\right)
}\equiv E_{in}\right\} $, i.e. the spectrum constructed by picking one level
from every $n$ levels in the original spectrum $\left\{ E_{i}\right\} $. By
this construction, the $n$-th order level spacing and gap ratio in $\left\{
E_{i}\right\} $ become the lowest-order ones in $\left\{ E_{i}^{\left(
n\right) }\right\} $. It will be verified that the joint probability
distribution of $\left\{ E_{i}^{\left( n\right) }\right\} $ (to leading
order) bears the same form as $\left\{ E_{i}\right\} $ with a rescaled
parameter $\gamma $ expressed in Eq.~(\ref{equ:0}), which hence explains
their distributions by virtue of Wigner surmise. Furthermore, this rescaling
relation holds for general $\beta $ beyond $\beta =1,2,4$, therefore, the
higher-order level spacing and gap ratio distributions in these ensembles
can be obtained accordingly, which is thus a natural extension of Ref.~[%
\onlinecite{Tekur}].

This paper is organized as follows. In Sec.~\ref{sec2} we summarize the
formulas regarding the higher-order level spacings and gap ratios, which
motivates the construction of reduced energy spectrum. In Sec.~\ref{sec3} we
focus on the cases that $\left\{ E_{i}^{\left( n\right) }\right\} $ has two
and three levels, and provide compelling numerical evidence for the scaling
of its joint probability distribution in ensembles with general $\beta $. In
Sec.~\ref{sec4} we provide numerical simulations for the distributions of
nearest level spacing and gap ratio in $\left\{ E_{i}^{\left( n\right)
}\right\} $ with large number of levels, and confirm their coincidence with
the higher-order ones in $\left\{ E_{i}\right\} $. In Sec.~\ref{sec5} we
briefly discuss the higher-order reduced energy spectrums. Conclusion and
discussion come in Sec.~\ref{sec6}.

\section{Motivating The Reduced Energy Spectrum}

\label{sec2}

The starting point to study the spectral statistics in the Gaussian ensemble
of random matrix is the joint probability distribution (JPDT) of its energy
levels, whose form is\cite{Mehta,Haake2001}%
\begin{equation}
P\left( \beta ,\left\{ E_{i}\right\} \right) =C\prod_{i<j}\left\vert
E_{i}-E_{j}\right\vert ^{\beta }e^{-A\sum_{i=1}^{N}E_{i}^{2}}\text{,}
\label{equ:Dist0}
\end{equation}%
where the Dyson index $\beta \in \left( 0,\infty \right) $ is a continuous
parameter, $C$ and $A$ are coefficients correlated by the normalization
condition $\int \prod_{i}dE_{i}P\left( \beta ,\left\{ E_{i}\right\} \right)
=1$. It's worth noting that only the subset $\beta =1,2,4$ are physically
invariant ensembles, that is, each matrix element is allowed to drawn from a
Gaussian distribution provided the matrix being invariant under orthogonal,
unitary or symplectic transformations. While for other values of $\beta $,
the JPDT stems from a special tridiagonal random matrix (see Eq.~(\ref%
{equ:Beta}) in Sec.~\ref{sec4}).

From the JPDT in Eq.~(\ref{equ:Dist0}), we can in principle calculate any
statistical quantity we want, in particular the distribution of nearest
level spacings and gap ratios. The general distributions for them in large
dimension $N$ are complicated, while for most practical purpose it is
sufficient to adopt the so-called Wigner surmise that deals with smallest
matrix that holds the quantity of interest. For example, to study nearest
level spacing it's sufficient to consider a $2\times 2$ matrix, which gives
the celebrated Wigner-Dyson distribution\cite{Haake2001}
\begin{equation}
P\left( \beta ,s\right) =C\left( \beta \right) s^{\beta }e^{-A\left( \beta
\right) s^{2}}\text{,}  \label{equ:WD}
\end{equation}%
where the coefficients $C\left( \beta \right) ,A\left( \beta \right) $ are
determined by the normalization conditions%
\begin{equation}
\int_{0}^{\infty }P\left( \beta ,s\right) ds=1\text{, }\int_{0}^{\infty
}sP\left( \beta ,s\right) ds=1\text{.}
\end{equation}%
It's easy to see $P\left( \beta ,s\rightarrow 0\right) \sim s^{\beta }$,
hence $\beta $ is the parameter that controls the strength of level
repulsion.

As for the nearest gap ratios $\left\{ r_{i}=s_{i+1}/s_{i}\right\} $, a
Wigner like surmise is applicable by studying $3\times 3$ matrices\cite{Atas}%
, which gives%
\begin{equation}
P\left( \beta ,r\right) =\frac{1}{Z_{\beta }}\frac{\left( r+r^{2}\right)
^{\beta }}{\left( 1+r+r^{2}\right) ^{1+3\beta /2}}  \label{equ:Pr}
\end{equation}%
where $Z_{\beta }$ is the normalization factor determined by requiring $%
\int_{0}^{\infty }P\left( \beta ,r\right) dr=1$. It is crucial to note that
the derivations for Eq.~(\ref{equ:WD}) and Eq.~(\ref{equ:Pr}) are purely
mathematical, that is, applicable for arbitrary positive $\beta $.

For the higher order level spacings $\left\{ s_{i}^{\left( n\right)
}=E_{i+n}-E_{i}\right\} $, its distribution is studied in Ref.~[%
\onlinecite{Rao20}] using a Wigner-like surmise that deals with $\left(
n+1\right) \times \left( n+1\right) $ matrix, and the result shows they
follow a generalized Wigner-Dyson distribution that bears the same form as
Eq.~(\ref{equ:WD}) with the parameter $\beta $ rescaled to $\gamma $ as
expressed in Eq.~(\ref{equ:0}).

On the other hand, higher order gap ratios come in two different ways, i.e.
the ``overlapping'' way\cite{Atas2} and ``non-overlapping'' way\cite{Tekur}.
In the former case we are dealing with%
\begin{equation}
\widetilde{r}_{i}^{\left( n\right) }=\frac{E_{i+n+1}-E_{i+1}}{E_{i+n}-E_{i}}=%
\frac{s_{i+n}+s_{i+n-1}+...+s_{i+1}}{s_{i+n-1}+s_{i+n-2}+...+s_{i}}\text{,}
\end{equation}%
which is termed ``overlapping'' gap ratio since there are shared spacings
between the numerator and denominator. While the $n$-th order
``non-overlapping'' gap ratio is defined as%
\begin{equation}
r_{i}^{\left( n\right) }=\frac{E_{i+2n}-E_{i+n}}{E_{i+n}-E_{i}}=\frac{%
s_{i+2n-1}+s_{i+2n-2}+...+s_{i+n}}{s_{i+n-1}+s_{i+n-2}+...+s_{i}}\text{.}
\end{equation}%
Both these two generalizations reduce to the nearest gap ratio when $n=1$,
but they are very different when studying their distributions using Wigner
surmise: for non-overlapping ratio $r^{\left( n\right) }$, the smallest
matrix dimension is $\left( 2n+1\right) \times \left( 2n+1\right) $; while
it is $\left( n+2\right) \times \left( n+2\right) $ for overlapping ratios.
Naively, it's expected that $P\left( \widetilde{r}^{\left( n\right) }\right)
$ is more complicated due to the overlapping spacings. Indeed, the
analytical form of $P\left( \widetilde{r}^{\left( 2\right) }\right) $ was
worked out in Ref.~[\onlinecite{Atas2}] and the result is quite involving.
For the non-overlapping ratios, Ref.~[\onlinecite{Tekur}] provides
compelling numerical evidence that in cases with $\beta =1,2,4$, $P\left(
r^{\left( n\right) }\right) $ bears the same form as $P\left( r\right) $
with the same rescaling parameter as higher order level spacing, that is,
Eq.~(\ref{equ:0}).

In summary, we have%
\begin{eqnarray}
P\left( \beta ,s^{\left( n\right) }\right) &=&P\left( \gamma ,s\right) \text{%
,}  \label{equ:Dists} \\
\text{\thinspace }P\left( \beta ,r^{\left( n\right) }\right) &=&P\left(
\gamma ,r\right) \text{,}  \label{equ:Dists2}
\end{eqnarray}%
where $\gamma $ is expressed in Eq.~(\ref{equ:0}). For the rest of this
paper, ``gap ratio'' will always refer to the non-overlapping one.

The identical rescaling behavior for higher level spacing and gap ratio
hints they may be attributed to one single reason, which is found to be the
reduced energy spectrum $\left\{ E_{i}^{\left( n\right) }\right\} $.
Formally, a reduced energy spectrum $\left\{ E_{i}^{\left( n\right)
}\right\} $ is constructed by picking one level from every $n$ levels in the
original spectrum $\left\{ E_{i}\right\} $, which is mathematically achieved
by tracing out every $n-1$ levels in between. This construction is very
similar to that of the reduced density matrix there we trace out the degrees
of freedom in a subsystem, hence $\left\{ E_{i}^{\left( n\right) }\right\} $
is named ``reduced energy spectrum''. By this construction, the $n$-th order
level spacing and gap ratio in $\left\{ E_{i}\right\} $ are mapped to the
lowest-order counterparts in $\left\{ E_{i}^{\left( n\right) }\right\} $.
It's then natural to conjecture that $\left\{ E_{i}\right\} $ and $\left\{
E_{i}^{\left( n\right) }\right\} $ (to leading order) bear the same form for
their probability distributions, with the Dyson index $\beta $ for the
former rescaled to $\gamma $ for the latter according to Eq.~(\ref{equ:0}).
Therefore, by applying Wigner surmise to $\left\{ E_{i}^{\left( n\right)
}\right\} $, the scaling behaviors in Eq.~(\ref{equ:Dists}) and Eq.~(\ref%
{equ:Dists2}) can be explained simultaneously. This is the main task of
current work.

Before proceeding, we want to mention the relatively trivial case of Poisson
ensemble. The reduced energy spectrum in Poisson ensemble has been studied
in Ref.~[\onlinecite{Daisy}] (which is named ``Daisy model'' by the
authors), where $n$-th order level spacing is shown to follow the
generalized semi-Poisson distribution
\begin{equation}
P\left( s^{\left( n\right) }=s\right) =\frac{n^{n}}{(n-1)!}s^{n-1}e^{-ns}%
\text{,}  \label{equ:P1}
\end{equation}%
which reduces to the conventional Poisson distribution $P\left( s\right)
=\exp \left( -s\right) $ when $n=1$. For the $n$-th order gap ratios, we can
derive their distribution from the results in Ref.~[\onlinecite{Daisy}],
that is
\begin{equation}
P\left( r^{(n)}=r\right) =\frac{r^{n-1}}{\left( 1+r\right) ^{2n}}\text{,}
\label{equ:P2}
\end{equation}%
which reduces to the one given in Ref.~[\onlinecite{Atas}] when $n=1$. The
formulas in Eq.~(\ref{equ:Dists}), Eq.~(\ref{equ:Dists2}), Eq.~(\ref{equ:P1}%
) and Eq.~(\ref{equ:P2}) will be used for later numerical simulations.

\section{Scaling of $P\left( \left\{ E_{i}^{\left( n\right) }\right\}
\right) $}

\label{sec3}

\bigskip By the construction of reduced energy spectrum $\left\{
E_{i}^{\left( n\right) }\right\} $, its formal joint probability
distribution is%
\begin{equation}
P\left( \left\{ E_{i}^{\left( n\right) }\right\} \right)
=\prod_{i}\int_{E_{in}}^{E_{\left( i+1\right) n}}\prod_{j=in+1}^{\left(
i+1\right) n-1}dE_{j}P\left( \beta ,\left\{ E_{i}\right\} \right) \text{.}
\label{equ:Dist}
\end{equation}%
For reasons described in previous section, we conjecture $P\left( \left\{
E_{i}^{\left( n\right) }\right\} \right) $ (to leading order) bear the same
form as $P\left( \left\{ E_{i}\right\} \right) $ with the rescaled parameter
$\gamma $ as in Eq.~(\ref{equ:0}), that is,%
\begin{equation}
P\left( \left\{ E_{i}^{\left( n\right) }\right\} \right) \sim
\prod_{i<j}\left\vert E_{i}^{\left( n\right) }-E_{j}^{\left( n\right)
}\right\vert ^{\gamma }e^{-A^{\prime }\sum_{i=1}^{N/n}\left( E_{i}^{\left(
n\right) }\right) ^{2}}\text{.}  \label{equ:Rescale}
\end{equation}%
An analytical derivation from Eq.~(\ref{equ:Dist}) to Eq.~(\ref{equ:Rescale}%
) for arbitrary matrix dimension $N$ is mathematically challenging, only the
special case with $\beta =2/k$ ($k$ being positive integer) has been proven
rigorously in Ref.~[\onlinecite{Forrester}]. It is not our purpose to give a
general proof, instead, to explain the behaviors of the $P\left( s^{\left(
n\right) }\right) $ and $P\left( r^{\left( n\right) }\right) $ in Eq.~(\ref%
{equ:Dists}) and Eq.~(\ref{equ:Dists2}), we only need to verify Eq.~(\ref%
{equ:Rescale}) in the sense of Wigner surmise, that is, in the cases that $%
\left\{ E_{i}^{\left( n\right) }\right\} $ has only two and three levels,
for which we will provide strong numerical evidence in the following.

First of all, the constant $A^{\prime }$ is not important since it is only a
decay rate parameter, whose value can be tuned by global rescaling of the
energy levels without affecting the distribution of level spacing or gap
ratio. Therefore, we will focus on the parameter $\gamma $ that controls the
strength of level repulsion.

The ensembles with general positive $\beta $ can be divided into four
typical categories: (i) $\beta =1,2,4$, corresponding to the three standard
Gaussian ensembles, which are of most physical interest; (ii) $\beta$ is an
integer that goes beyond the three standard ensembles, for which we choose $%
\beta =3$; (iii) $\beta$ is a fraction, for which we choose $\beta =1/3$;
(iv) $\beta$ is an irrational value, for which we choose $\beta =\left(
\sqrt{5}-1\right) /2$ (the golden ratio). We will verify the recaling
relation Eq.~(\ref{equ:Rescale}) in these cases.

\begin{figure*}[htbp]
\centering
\includegraphics[width=2\columnwidth]{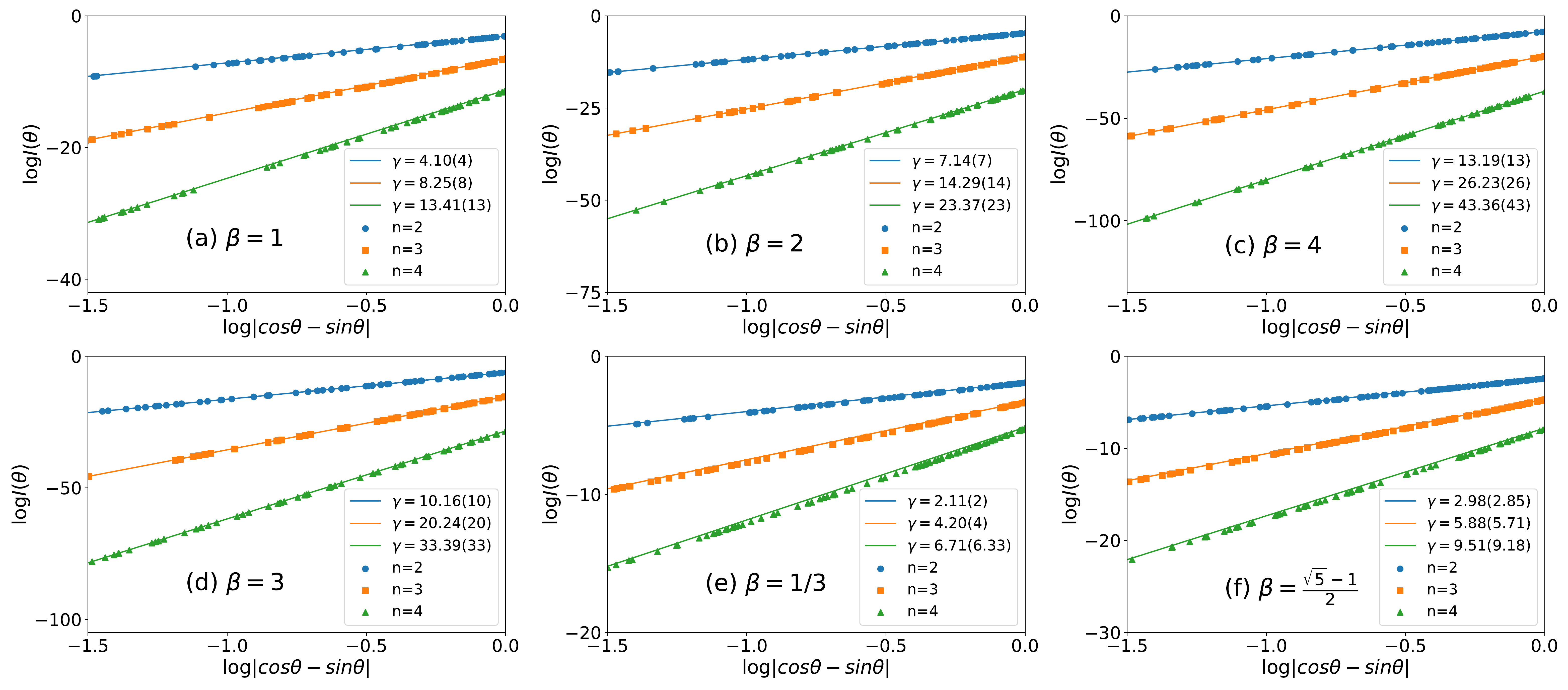}
\caption{The fitting of Eq.~(\protect\ref{equ:1}) for $\protect\beta=1,2,3,4,%
\frac{1}{3},\frac{\protect\sqrt{5}-1}{2}$ with $n=2,3,4$. The fitted slopes
are shown in the figure legends, where the numbers in the bracket are the
expected values according to Eq.~(\protect\ref{equ:0}). }
\label{fig:rescale0}
\end{figure*}

We start with the case that $\left\{ E_{i}^{\left( n\right) }\right\} $ has
only two levels, where the rescaling in Eq.~(\ref{equ:Rescale}) becomes%
\begin{eqnarray}
I\left( E_{0},E_{n}\right)
&=&\int_{E_{0}}^{E_{n}}\prod_{i=1}^{n-1}dE_{i}P\left( \beta ,\left\{
E_{i}\right\} \right) \\
&\sim &\left\vert E_{0}-E_{n}\right\vert ^{\gamma }e^{-A^{\prime }\left(
E_{0}^{2}+E_{n}^{2}\right) }\text{.}  \notag
\end{eqnarray}%
Denote $E_{0}=R\cos \theta $ and $E_{n}=R\sin \theta $, and keeping $R$
constant, we reach to%
\begin{equation}
\log I\left( \theta \right) =\gamma \log \left\vert \cos \theta -\sin \theta
\right\vert +\text{const.}  \label{equ:1}
\end{equation}%
Without loss of generality, we take $A=1$ and $R=1$, and randomly generate $%
200$ sets of $\theta $ in the range $[0,2\pi )$. We then numerically
calculate $\log I\left( \theta \right) $ and $\log \left\vert \cos \theta
-\sin \theta \right\vert $, and collect the results for $n=2,3,4$, which are
presented in Fig.~\ref{fig:rescale0}. As can be seen, the $\log I\left(
\theta \right) $ and $\log \left\vert \cos \theta -\sin \theta \right\vert $
shows a perfect linear dependence in all cases, with the fitted values of $%
\gamma $ quite close to the expected ones in Eq.~(\ref{equ:0}), as
summarized in Table~\ref{tab:1}.

\begin{figure*}[htbp]
\centering
\includegraphics[width=2\columnwidth]{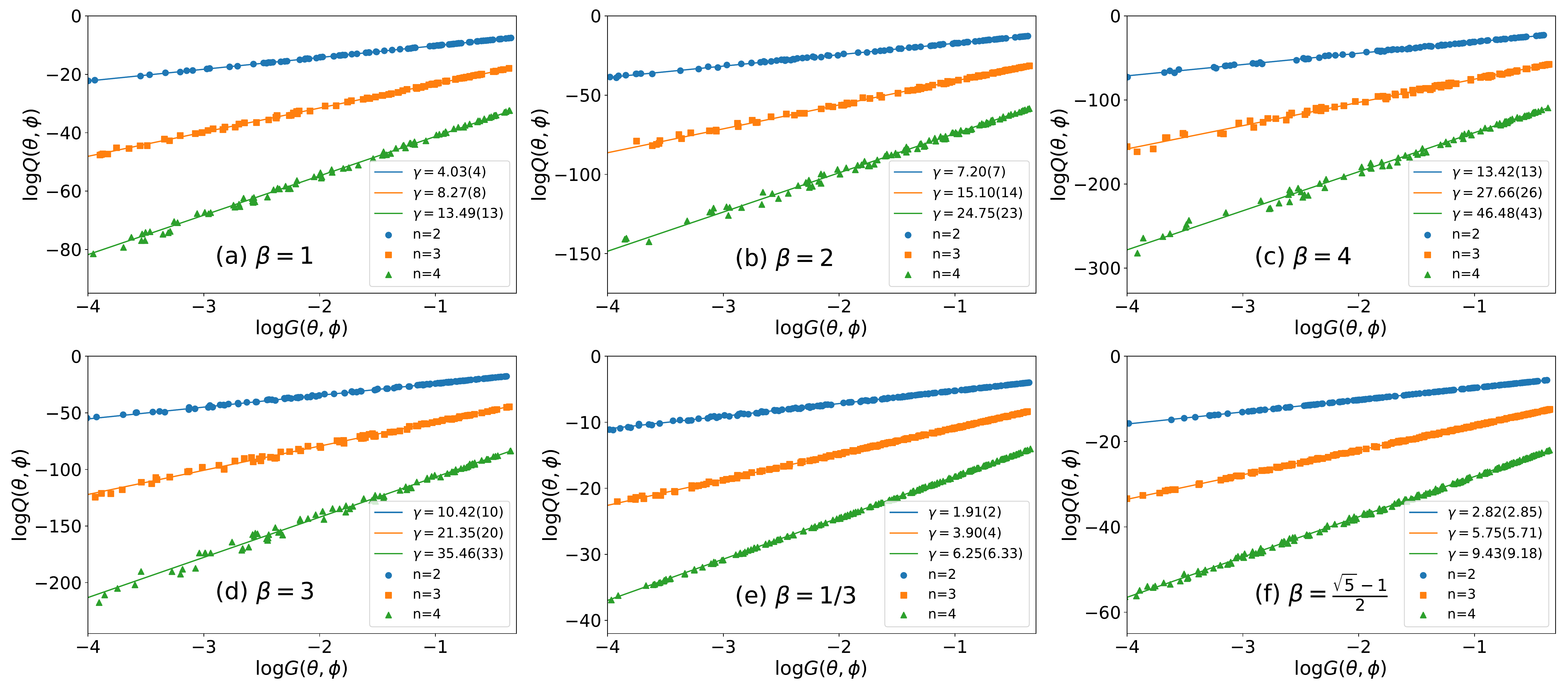}
\caption{The fitting of Eq.~(\protect\ref{equ:2}) for $\protect\beta=1,2,3,4,%
\frac{1}{3},\frac{\protect\sqrt{5}-1}{2}$ with $n=2,3,4$. The fitted $%
\protect\gamma$ values are shown in the figure legends, with the expected
values according to Eq.~(\protect\ref{equ:0}) in the brackets.}
\label{fig:rescale}
\end{figure*}

Next we consider the case that $\left\{ E_{i}^{\left( n\right) }\right\} $
has three levels, now the rescaling in Eq.~(\ref{equ:Rescale}) becomes
\begin{eqnarray}
&&Q\left( E_{0},E_{n},E_{2n}\right)  \notag \\
&\equiv
&\int_{E_{0}}^{E_{n}}\prod_{i=1}^{n-1}dE_{i}\int_{En}^{E_{2n}}%
\prod_{j=n+1}^{2n-1}dE_{j}P\left( \beta ,\left\{ E_{i}\right\} \right) \\
&\sim &\left\vert E_{0}-E_{n}\right\vert ^{\gamma }\left\vert
E_{n}-E_{2n}\right\vert ^{\gamma }\left\vert E_{0}-E_{2n}\right\vert
^{\gamma }e^{-A^{\prime }\left( E_{0}^{2}+E_{n}^{2}+E_{2n}^{2}\right) }\text{%
.}  \notag
\end{eqnarray}%
With the transformation to spherical coordinates%
\begin{equation}
E_{0}=R\sin \theta \cos \varphi ,E_{n}=R\sin \theta \sin \varphi
,E_{2n}=R\cos \theta
\end{equation}%
and keeping $R$ constant, we can reach to%
\begin{equation}
\log Q\left( R,\theta ,\varphi \right) =\gamma \log G\left( \theta ,\varphi
\right) +\text{const.}  \label{equ:2}
\end{equation}%
where%
\begin{eqnarray}
G\left( \theta ,\varphi \right) &=&|\left( \sin \theta \cos \varphi -\sin
\theta \sin \varphi \right)  \notag \\
&&\times \left( \sin \theta \cos \varphi -\cos \theta \right)  \notag \\
&&\times \left( \sin \theta \sin \varphi -\cos \theta \right) |\text{.}
\end{eqnarray}%
We then perform numerical checks, where we fix $R=1$ and $A=1$ as before. We
randomly generate $200$ pairs of $\left( \theta ,\varphi \right) $ and
numerically determine $\log Q\left( \theta ,\varphi \right) \equiv \log
Q\left( 1,\theta ,\varphi \right) $ and $\log G\left( \theta ,\varphi
\right) $, the results are displayed in Fig.~\ref{fig:rescale}. As can be
seen, the linear dependence between $\log Q\left( \theta ,\varphi \right) $
and $\log G\left( \theta ,\varphi \right) $ are still perfect in all cases,
and the fitted values of $\gamma $ are close to the expected ones in Eq.~(%
\ref{equ:0}).

For convenience, we collected the theoretical and numerical values of $%
\gamma $ in Table~\ref{tab:1}, where $\gamma _{\text{e}}$ refers to the
expected value according to Eq.~(\ref{equ:0}), $\gamma _{\text{num.}%
}^{\left( 1\right) }$ and $\gamma _{\text{num.}}^{\left( 2\right) }$ are the
numerical values fitted from Eq.~(\ref{equ:1}) and Eq.~(\ref{equ:2}), and
error$^{\left( 1\right) }$ and error$^{\left( 2\right) }$ are their relative
deviations from $\gamma _{\text{e}}$ respectively. In general, the numerical
errors increases with larger $\beta $, but they're controlled within a
satisfactory level in all cases.

\begin{table}
\centering%
\begin{tabular}{|c|c|c|c|c|c|c|}
\hline
$\beta $ & $n$ & $\gamma _{\text{e}}$ & $\gamma _{\text{num.}}^{\left(
1\right) }$ & error$^{\left( 1\right) }$ & $\gamma _{\text{num.}}^{\left(
2\right) }$ & error$^{\left( 2\right) }$ \\ \hline
& $2$ & $4$ & $4.10$ & $2.50\%$ & $4.03$ & $0.75\%$ \\ \cline{2-7}
$1$ & $3$ & $8$ & $8.25$ & $3.13\%$ & $8.27$ & $3.38\%$ \\ \cline{2-7}
& $4$ & $13$ & $13.41$ & $3.15\%$ & $13.49$ & $3.77\%$ \\ \hline
& $2$ & $7$ & $7.14$ & $2.00\%$ & $7.20$ & $2.86\%$ \\ \cline{2-7}
$2$ & $3$ & $14$ & $14.29$ & $2.07\%$ & $15.10$ & $7.86\%$ \\ \cline{2-7}
& $4$ & $23$ & $23.37$ & $1.61\%$ & $24.75$ & $7.29\%$ \\ \hline
& $2$ & $13$ & $13.19$ & $1.46\%$ & $13.42$ & $3.23\%$ \\ \cline{2-7}
$4$ & $3$ & $26$ & $26.23$ & $0.88\%$ & $27.66$ & $6.38\%$ \\ \cline{2-7}
& $4$ & $43$ & $43.36$ & $0.84\%$ & $46.48$ & $8.09\%$ \\ \hline
& $2$ & $10$ & $10.16$ & $1.60\%$ & $10.42$ & $4.20\%$ \\ \cline{2-7}
$3$ & $3$ & $20$ & $20.24$ & $1.20\%$ & $21.35$ & $6.75\%$ \\ \cline{2-7}
& $4$ & $33$ & $33.39$ & $1.18\%$ & $35.46$ & $7.45\%$ \\ \hline
& $2$ & $2$ & $2.11$ & $5.5\%$ & $1.91$ & $-4.5\%$ \\ \cline{2-7}
$1/3$ & $3$ & $4$ & $4.20$ & $5.0\%$ & $3.90$ & $-2.5\%$ \\ \cline{2-7}
& $4$ & $19/3$ & $6.71$ & $6.0\%$ & $6.25$ & $-1.3\%$ \\ \hline
& $2$ & $\frac{3\sqrt{5}-1}{2}$ & $2.98$ & $4.4\%$ & $2.82$ & $-1.2\%$ \\
\cline{2-7}
$\frac{\sqrt{5}-1}{2}$ & $3$ & $3\sqrt{5}-1$ & $5.88$ & $3.0\%$ & $5.75$ & $%
0.7\%$ \\ \cline{2-7}
& $4$ & $5\sqrt{5}-2$ & $9.51$ & $3.6\%$ & $9.43$ & $2.7\%$ \\ \hline
\end{tabular}
\caption{The values of $\gamma$ with different $\beta$ and order $n$,
where $\gamma_{\text{e}}$ is the expected value according to Eq.~(\ref{equ:0}).
$\gamma _{\text{num.}}^{\left( 1\right)}$ and $\gamma _{\text{num.}}^{\left( 2\right) }$
are the values fitted from Eq.~(\ref{equ:1}) and Eq.~(\ref{equ:2}),
their relative errors with respect to $\gamma_{\text{e}}$ are denoted by error$^{\left( 1\right) }$
and error$^{\left( 2\right) }$ respectively.} \label{tab:1}
\end{table}

Up to now, we have verified the scaling behavior for the probability
distribution of $\left\{ E_{i}^{\left( n\right) }\right\} $ with two and
three levels, the cases with more levels can be verified with the same
method, but it will become more and more tedious when the number of
integrals increases. Nevertheless, present results are sufficient to justify
the scaling behaviors of higher order level spacing/gap ratio by applying
Wigner surmise to $\left\{ E_{i}^{\left( n\right) }\right\} $. More
importantly, we have verified the scaling behavior to hold for general $%
\beta $s that go beyond GOE,GUE,GSE ($\beta =1,2,4$), even when $\beta $ is
non-integer or irrational. This indicates the distribution for higher order
level spacing and gap ratio in Eq.~(\ref{equ:Dists}) and Eq.~(\ref%
{equ:Dists2}) also hold for general $\beta $ ensembles, for which we will
present numerical evidences in the next section.

\section{Numerical Simulations}

\label{sec4}

\bigskip In this section, we numerically check the distribution of nearest
level spacing/gap ratio in $\left\{ E_{i}^{\left( n\right) }\right\} $ with
many levels, and show they indeed coincide with the $n$-th order
counterparts in $\left\{ E_{i}\right\} $. Before that, a technical issue
needs to be pointed out, that is, the nearest level spacings in $\left\{
E_{i}^{\left( n\right) }\right\} $ is actually $\left\{ E_{i+1}^{\left(
n\right) }-E_{i}^{\left( n\right) }=E_{\left( i+1\right) n}-E_{in}\right\} $%
, whose total number is $\left[ \frac{N}{n}\right] -1$; while the $n$-th
order level spacing in original energy spectrum $\left\{ E_{i}\right\} $ are
$\left\{ E_{i+n}-E_{i}\right\} $ with total number $N-n$; therefore the
mapping does not strictly hold, the same thing happens to gap ratios.
However, since the distribution is extracted from a large number of level
spacings (gap ratios), it's natural to suspect the difference is negligible
when the number of samples and matrix dimension are large, which we will
soon justify.

\begin{figure*}[htbp]
\centering
\includegraphics[width=2\columnwidth]{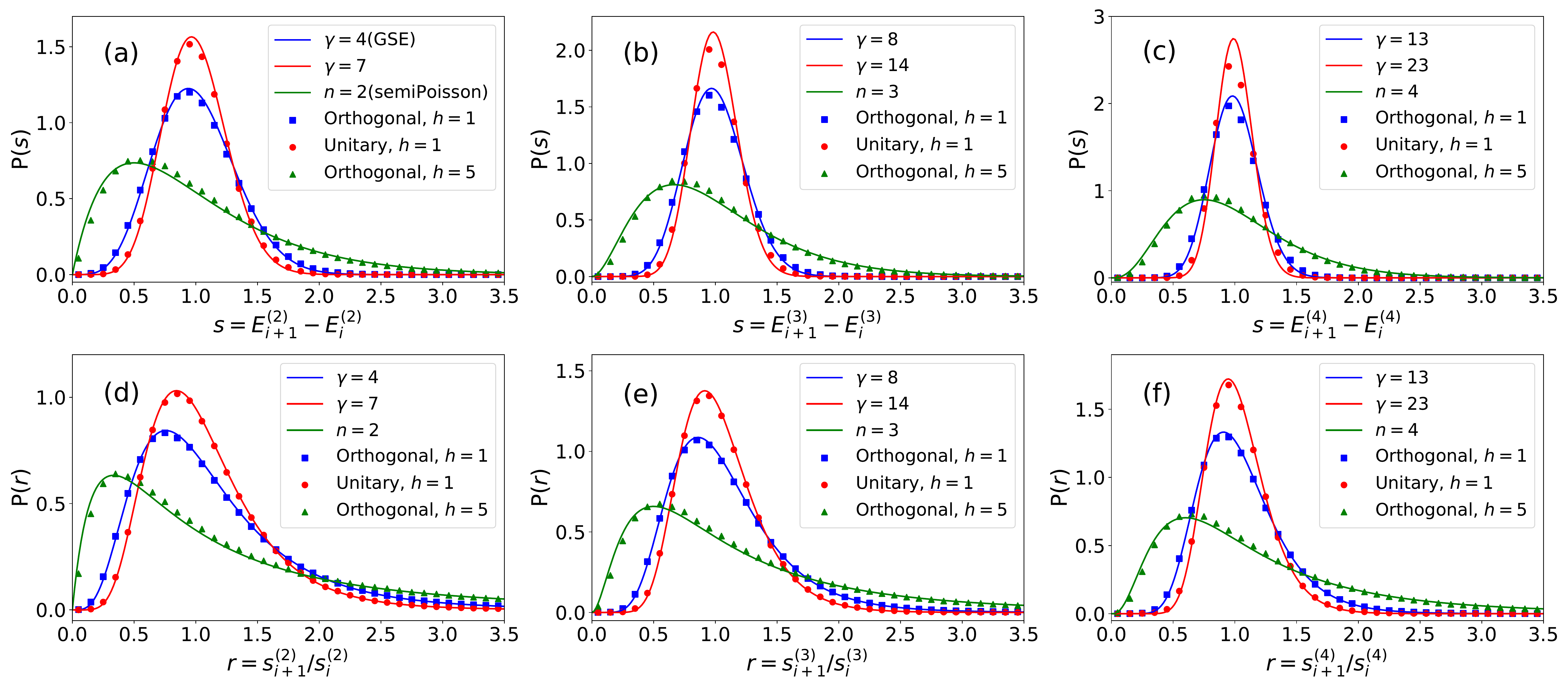}
\caption{Distribution of nearest level spacing and gap ratio in the reduced
energy spectrum $\left\{ E_{i}^{\left( n\right) }\right\}$ of model Eq.~(%
\protect\ref{equ:H}) in a $L=12$ chain with $n=2$ ((a) and (d)), $n=3$ ((b)
and (e)) and $n=4$((c) and (f)). The data from $h=1$ in the orthogonal
(unitary) model represent GOE (GUE), and those from $h=5$ in orthogonal
model represent Poisson respectively. The reference curves corresponds to
the higher-order spacing distributions in $\left\{E_{i}\right\}$ according
to Eq.~(\protect\ref{equ:Dists}), Eq.~(\protect\ref{equ:Dists2}) for GOE and
GUE, and Eq.~(\protect\ref{equ:P1}), Eq.~(\protect\ref{equ:P2}) for Poisson,
the parameter $\protect\gamma$ for the former is calculated by Eq.~(\protect
\ref{equ:0}). The perfect fittings in all cases confirms the coincidence
between higher-order spacing distributions in $\left\{E_{i}\right\}$ and the
lowest-order counterparts in $\left\{E_{i}^{(n)}\right\}$.}
\label{fig:Spin}
\end{figure*}

\bigskip For GOE,GUE and Poisson ensemble, we perform simulations from a
real physical system, that is, the 1D Heisenberg chain with random external
fields, which is the canonical model to study many-body localization\cite%
{Alet}. The Hamiltonian reads%
\begin{equation}
H=J\sum_{i=1}^{L}\mathbf{S}_{i}\cdot \mathbf{S}_{i+1}+\sum_{i=1}^{L}\sum_{%
\alpha =x,y,z}h^{\alpha }\varepsilon _{i}^{\alpha }S_{i}^{\alpha }\text{,}
\label{equ:H}
\end{equation}%
where $\mathbf{S}_{i}$ is spin-$1/2$ operator. The anti-ferromagnetic
coupling strength $J$ is set to be unity, and $\varepsilon _{i}^{\alpha }$s
are random numbers within range $\left[ -1,1\right] $. The $h^{\alpha }$ is
referred as randomness strength. We consider two sets of $h^{\alpha }$: (i) $%
h^{x}=h^{z}=h\neq 0$ and $h^{y}=0$, the model is orthogonal and belongs to
GOE; (ii) $h^{x}=h^{y}=h^{z}=h\neq 0$, the model is unitary and belongs to
GUE. This model undergos a thermal-MBL transition at roughly $h_{c}\simeq 3$
($2.5$) in the orthogonal (unitary) case, where the level spacing
distribution evolves from GOE (GUE) to Poisson\cite{Regnault16,Regnault162}.

We choose a $L=12$ system to perform simulations, and prepare $500$ samples
of energy spectrum at $h=1$ in both the orthogonal and unitary model,
representing GOE and GUE respectively. We also prepare $500$ samples at $h=5$
in the orthogonal model to represent Poisson ensemble. For each sample of
energy spectrum, we manually construct the reduced energy spectrum $\left\{
E_{i}^{\left( n\right) }\right\} $ with $n=2,3,4$ and count the
corresponding nearest level spacing and gap ratio distributions, and compare
them to the formulas in Eq.~(\ref{equ:Dists})-(\ref{equ:P2}), the results
are displayed in Fig.~\ref{fig:Spin}. As can be seen, the fittings are quite
good, confirming the correspondence between nearest level spacing/gap ratio
in $\left\{ E_{i}^{\left( n\right) }\right\} $ with the $n$-th order
counterparts in $\left\{ E_{i}\right\} $.

For ensembles with general $\beta $, we perform numerical simulations from
modelling random matrices. It was proven in Ref.~[\onlinecite{Beta}] that
the eigenvalues of the following tridiagonal matrix ensemble%
\begin{equation}
M_{\beta }=\frac{1}{\sqrt{2}}\left(
\begin{array}{ccccc}
x_{1} & y_{1} &  &  &  \\
y_{1} & x_{2} & y_{2} &  &  \\
&
\begin{array}{ccc}
\text{.} &  &  \\
& \text{.} &  \\
&  & \text{.}%
\end{array}
&
\begin{array}{ccc}
\text{.} &  &  \\
& \text{.} &  \\
&  & \text{.}%
\end{array}
&
\begin{array}{ccc}
\text{.} &  &  \\
& \text{.} &  \\
&  & \text{.}%
\end{array}
&  \\
&  & y_{N-2} & x_{N-1} & y_{N-1} \\
&  &  & y_{N-1} & x_{N}%
\end{array}%
\right)  \label{equ:Beta}
\end{equation}%
will follow the distribution in Eq.~(\ref{equ:Dist0}) with continuous
parameter $\beta \in \left( 0,\infty \right) $ provided the diagonals $%
x_{i}\,$($i=1,2,...,N$) follow the normal distribution $\mathit{N}\left(
0,2\right) $, that is%
\begin{equation}
P\left( x_{i}\right) =\frac{1}{2\sqrt{2\pi }}e^{-x_{i}^{2}/8}\text{, }%
i=1,2,...,N\text{,}
\end{equation}%
and $y_{k}$ ($k=1,2,...,N-1$) follows the $\chi $ distribution with
parameter $\left( N-k\right) \beta $, that is%
\begin{equation}
P\left( y_{k}\right) =\left\{
\begin{array}{ll}
\frac{2}{2^{\left( N-k\right) \beta }\Gamma \left( \left( N-k\right) \beta
/2\right) }y_{k}^{\left( N-k\right) \beta -1}e^{-y_{k}^{2}/2}\text{,} &
y_{k}\geq 0 \\[1mm]
0\text{,} & y_{k}<0%
\end{array}%
\right. \text{.}
\end{equation}%
By virtue this remarkable construction, we can efficiently generate energy
spectrums with any $\beta $ we want\cite{note}.

\begin{figure*}[tbph]
\centering
\includegraphics[width=2\columnwidth]{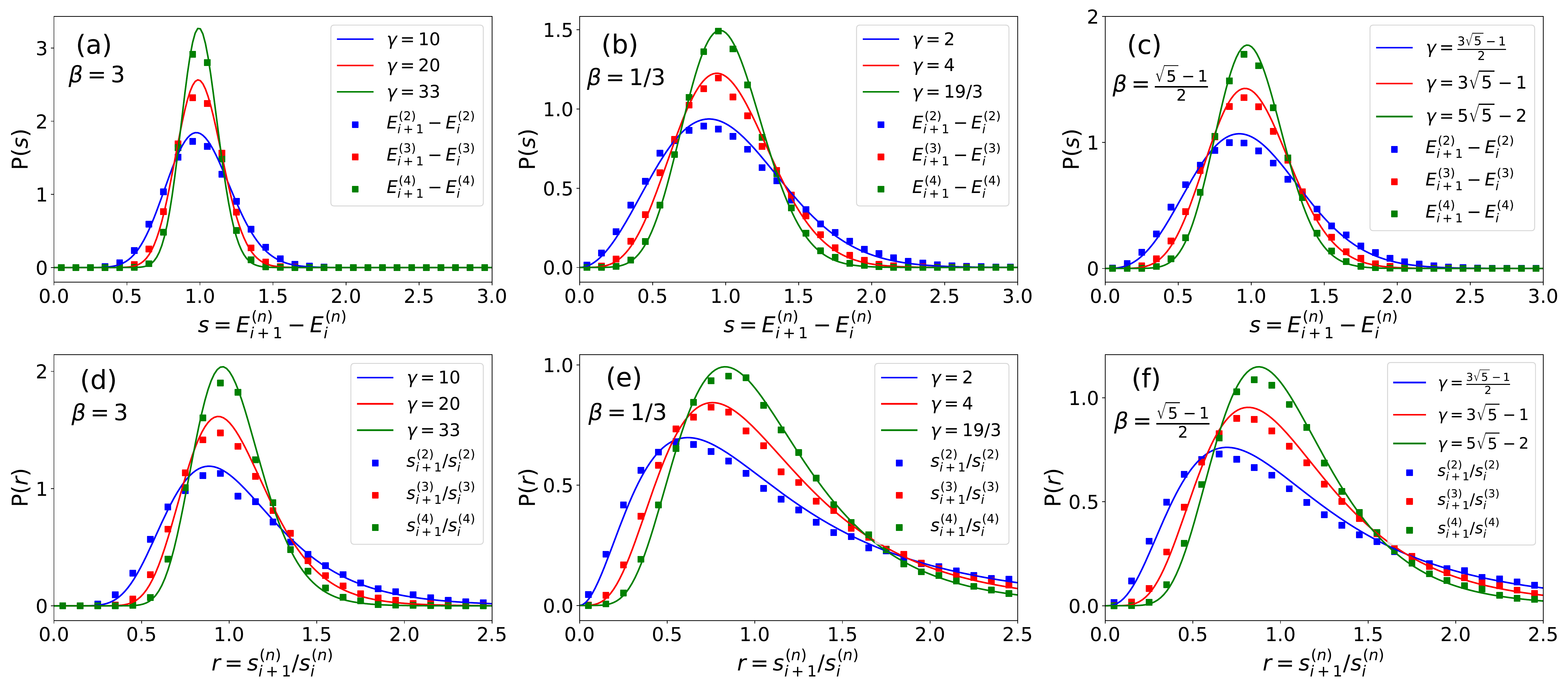}
\caption{Distribution of nearest level spacing and gap ratio in the reduced
energy spectrum $\left\{ E_{i}^{\left( n\right) }\right\}$ with $n=2,3,4$
from ensembles with $\protect\beta=3$ ((a),(d)), $\frac{1}{3}$ ((b),(e)) and
$\frac{\protect\sqrt{5}-1}{2}$ ((c),(f)), and the reference curves with
index $\protect\gamma$ correspond to the ones in Eq.~(\protect\ref{equ:Dists}%
) and Eq.~(\protect\ref{equ:Dists2}) for level spacing and gap ratio
respectively. These fittings confirm the higher-order spacing distributions
in general $\protect\beta$ ensembles.}
\label{fig:beta}
\end{figure*}

In accordance with Sec.~\ref{sec3}, we choose to simulate the cases with $%
\beta =3,\frac{1}{3},\frac{\sqrt{5}-1}{2}$. For each $\beta $, we generate $%
500$ samples of energy spectrums by using Eq.~(\ref{equ:Beta}), with the
number of energy levels in the original spectrum $\left\{ E_{i}\right\} $
kept to be $600$. Then we construct the corresponding reduced energy
spectrums $\left\{ E_{i}^{\left( n\right) }\right\} $ with $n=2,3,4$ and
determine the distributions of nearest level spacings/gap ratios in them,
the results are shown in Fig.~\ref{fig:beta}. As can be seen, the fittings
are quite satisfactory in all cases.

Up to now, we have verified the coincidence between nearest level
spacing/gap ratio in $\left\{ E_{i}^{\left( n\right) }\right\} $ with the $n$%
-th order ones in $\left\{ E_{i}\right\} $\ for general Gaussian $\beta $
ensembles. As a result, we have generalized the scaling behavior for $%
P\left( s^{\left( n\right) }\right) $ and $P\left( r^{\left( n\right)
}\right) $ in Eq.~(\ref{equ:Dists}) and Eq.~(\ref{equ:Dists2}) to cases
beyond GOE,GUE,GSE. The Gaussian ensemble with non-integer $\beta $ has been
used to describe the spacing distributions along the thermal-MBL transition%
\cite{Buijsman}, while its efficiency in describing long-range level
correlations is under debate\cite{Sierant19}. Our results thus provide a
numerical criteria for such studies.

\section{Higher-order Reduced Energy Spectrums}

\label{sec5}

\bigskip Results in Sec.~\ref{sec3} verified the scaling of $\left\{
E_{i}^{\left( n\right) }\right\} $ in cases with two/three levels, which is
sufficient for applying Wigner surmise to $\left\{ E_{i}^{\left( n\right)
}\right\} $ regarding level spacing/gap ratio, and the numerical simulations
in Sec.~\ref{sec4} provide strong support for the cases where $\left\{
E_{i}^{\left( n\right) }\right\} $ has more levels. Actually, we can
continue to construct the $2$nd order reduced energy spectrums, i.e. the
``reduced energy spectrum of reduced energy spectrum'', whose nearest level
spacing/gap ratio will correspond to the higher-order ones in $\left\{
E_{i}^{\left( n\right) }\right\} $, and will rescale in a similar manner.
Denote the $m$-th order level spacing and gap ratio in $\left\{
E_{i}^{\left( n\right) }\right\} $ as $s_{i}^{\left( n,m\right) }$ and $%
r_{i}^{\left( n,m\right) }$, that is,%
\begin{equation}
s_{i}^{\left( n,m\right) }=E_{i+m}^{\left( n\right) }-E_{i}^{\left( n\right)
}\text{, }r_{i}^{\left( n,m\right) }=\frac{E_{i+2m}^{\left( n\right)
}-E_{i+m}^{\left( n\right) }}{E_{i+m}^{\left( n\right) }-E_{i}^{\left(
n\right) }}\text{,}
\end{equation}%
it's straightforward to write down their expected probability distributions%
\begin{eqnarray}
&&P\left( \beta ,s^{\left( n,m\right) }\right) =P\left( \delta ,s\right)
\text{, }P\left( \beta ,r^{\left( n,m\right) }\right) =P\left( \delta
,r\right) \\
&&\delta =\frac{m(m+1)}{2}\gamma +m-1\text{, }\gamma =\frac{n(n+1)}{2}\beta
+n-1\text{.}  \notag
\end{eqnarray}%
The same procedure can continue for higher-order reduced energy spectrums.

Of course, such a construction is artificial, meanwhile, it reveals a
hierarchy of energy spectrums can emerge from single spectrum, which (to
lowest order) bear the same form of probability distributions. Moreover, by
taking a closer look at the scaling expression $\gamma =\frac{n\left(
n+1\right) }{2}\beta +n-1$, we immediately recognize an infinite number of
coincident relations between $\left\{ E_{i}^{\left( n\right) }\right\} $
from different ensembles. For examples, $\left\{ E_{i}^{\left( 2\right)
}\right\} $ in $\beta =1$ (GOE) has the same structure as $\left\{
E_{i}\right\} $ in $\beta =4$ (GSE) -- a result known before\cite%
{GSE,Forrester}, and both of them coincide with $\left\{ E_{i}^{\left(
3\right) }\right\} $ in $\beta =\frac{1}{3}$. Actually, it's easy to verify
that for $\gamma \in (k,k+1]$ there exists $k$ different sets of $\left(
\beta ,n\right) $ that have equal $\gamma $, and their lower-order level
statistics are expected to be identical.

\section{Conclusion and Discussion}

\label{sec6}

We studied the reduced energy spectrum $\left\{ E_{i}^{\left( n\right)
}\right\} $, constructed by picking one level from every $n$ levels in
original spectrum $\left\{ E_{i}\right\} $. It is verified the distribution
of $\left\{ E_{i}^{\left( n\right) }\right\} $ (to leading order) bears the
same form as $\left\{ E_{i}\right\} $, with the Dyson index rescaled from $%
\beta $ to $\gamma =\frac{n\left( n+1\right) }{2}\beta +n-1$. It's then
demonstrated the nearest level spacing and gap ratio in $\left\{
E_{i}^{\left( n\right) }\right\} $ correspond to the $n$-th order ones in $%
\left\{ E_{i}\right\} $, which explains the distributions of the latter
found recently in Ref.~[\onlinecite{Rao20}] and Ref.~[\onlinecite{Tekur}]
simultaneously.

Moreover, we find the rescaling of reduced energy spectrum holds for
Gaussian ensembles that go beyond the standard GOE,GUE,GSE, and establish
the distributions of higher-order level spacings and gap ratios in these
ensembles. We also confirmed such correspondences in the Poisson ensemble,
and discovered the distribution of $n$-th order gap ratios, as expressed in
Eq.~(\ref{equ:P2}).

It's noted the reduced energy spectrum has been studied for the Poisson
ensemble\cite{Daisy} and some Gaussian ensembles, that is, the special ones
with $\beta =2/k$ ($k$ being positive integer)\cite{Forrester}, which
contains the well-known coincidence between $\left\{ E_{i}^{\left( 2\right)
}\right\} $ in GOE and $\left\{ E_{i}\right\} $ in GSE. Our work is thus a
natural extension of these studies.

\bigskip The significance of our work is three-folded. First, we explained
the distributions of higher-order level spacings and gap ratios -- both of
which are widely used in the study of MBL and whose distributions are found
separately in recent studies -- by a single common mechanism: the reduced
energy spectrum. Second, we generalized the higher-order spacing
distributions in Eq.~(\ref{equ:Dists}) and Eq.~(\ref{equ:Dists2}) to general
$\beta $ ensembles, which may be beneficial for studying systems that go
beyond the standard Gaussian ensembles\cite{Lucas,Tzortzakakis,Buijsman,Ahmed,Jain2}. Third, our results reveals a rich set of
structures hidden in the energy spectrum, by constructing the reduced energy
spectrums.

\bigskip Last but not least, in our numerical simulations in Sec.~\ref{sec4}%
, we were employing the modelling matrix\ of general $\beta $ ensemble as
expressed in Eq.~(\ref{equ:Beta}). It's thus natural and interesting to ask
whether this ``parent matrix'' corresponds to a real quantum system, and
what's the property of such a system if it does. Given the physically
relevant GOE,GUE,GSE are incorporated in Eq.~(\ref{equ:Beta}), we conjecture
such a ``parent Hamiltonian'' does exist, whose construction is left for a
future study.

\section*{Acknowledgements}

This work is supported by the National Natural Science Foundation of China
through Grant No.11904069 and No.11847005 and No.11804070.


\begin{thebibliography}{99}
\bibitem{note} For standard GOE,GUE and GSE, we can model either by their
physical construction or by using Eq.~(\ref{equ:Beta}) with $\beta=1,2,4$ to
get the energy spectrums, although the DOS might be different. We have
verified that these two constructions work almost equally well when studying
spacing distributions, even when the matrix is as small as $200\times 200$.
\end{thebibliography}
\end{document}